\documentclass[prl,aps,reprint,showpacs,floatfix]{revtex4-1}
\usepackage{mathrsfs}
\usepackage{color}
\usepackage{graphicx}           
\usepackage{dcolumn}            
\usepackage{bm}                 
\usepackage{hyperref}           
\usepackage[latin1]{inputenc}   
\usepackage{pstricks}           
\usepackage{amsmath,amssymb}
\usepackage{version}
\usepackage{epstopdf}
\usepackage{ulem}
\usepackage{upgreek}

\DeclareGraphicsExtensions{.eps,.ps}

\newcommand{\grad}{\ensuremath{^{\circ}}}

\begin{document}

\title{Disentangling the role of laser coupling in directional breaking of molecules}

\author{Qiying Song$^1$, Zhichao Li$^2$, Sen Cui$^2$, Peifen Lu$^1$, Xiaochun Gong$^1$, Qinying Ji$^1$, Kang Lin$^1$, Wenbin Zhang$^1$, Junyang Ma$^1$, Haifeng Pan$^1$, Jingxin Ding$^1$,  Matthias F. Kling$^{3,4}$}
\author{Heping Zeng$^{1}$}
\email{Email: hpzeng@phy.ecnu.edu.cn}
\author{Feng He$^{2}$}
\email{Email: fhe@sjtu.edu.cn}
\author{Jian Wu$^{1,5}$}
\email{Email: jwu@phy.ecnu.edu.cn}

\affiliation{
$^1$State Key Laboratory of Precision Spectroscopy, East China Normal University, Shanghai 200062, China\\
 $^2$Key Laboratory of Laser Plasmas (Ministry of Education) and Department of Physics and Astronomy, Collaborative Innovation Center for IFSA (CICIFSA), Shanghai Jiao Tong University, Shanghai 200240, China\\
 $^3$Department of Physics, LMU Munich, Am Coulombwall 1, D-85748 Garching, Germany\\
 $^4$Max-Planck-Institut f\"{u}r Quantenoptik, Hans-Kopfermann-Stra$\beta$e 1, D-85748 Garching, Germany\\
 $^5$Collaborative Innovation Center of Extreme Optics, Shanxi University, Taiyuan 030006, China
}
\date{\today}

\begin{abstract}{
The directional control of molecular dissociation with a laser electric field waveform is a paradigm and was demonstrated for a variety of molecules. In most cases, the directional control occurs via a dissociative ionization pathway. The role of laser-induced coupling of electronic states in the dissociating ion versus selective ionization of oriented neutral molecules, however, could not be distinguished for even small heteronuclear molecules such as CO. Here, we introduce a technique, using elliptically polarized pump and linearly polarized two-color probe pulses, that unambiguously distinguishes the roles of laser-induced state coupling and selective ionization. The measured photoelectron momentum distributions governed by the light polarizations allow us to coincidentally identify the ionization and dissociation from the pump and probe pulses. Directional dissociation of CO$^+$ as a function of the relative phase of the linearly polarized two-color pulse is observed for both parallel and orthogonally oriented molecules. We find that the laser-induced coupling of various electronic states of CO$^+$ plays an important role for the observed directional bond breaking, which is verified by quantum calculations.
}
\end{abstract}
\maketitle

\section{INTRODUCTION}

As a primary step for steering chemical dynamics, directional bond breaking is one of the most fundamental and interesting phenomena in molecular dissociative ionization. It can be coherently controlled by using carrier-envelope phase (CEP) stabilized few-cycle \cite{1,2,3,4,5} or two-color \cite{6,7,8,9,10} ultrashort laser pulses. Previous studies have demonstrated that the directional bond breaking in dissociative ionization of a diatomic molecule with symmetric orbital profile along the molecular axis is governed by the pathway interference of the dissociating nuclear wave packets \cite{11,12,13,14,15}. The most intensively investigated example is the dissociative single ionization of molecular hydrogen and its isotopes \cite{1,2,3,4,5,6,8,11,12,13,14,15,16,17,18,19}, which have symmetric electron distribution along the molecular axis.

For many heteronuclear diatomic molecules, the electron distributions around two nuclei are asymmetric, thus both the ionization and dissociation may be directional and contribute to the ultimate asymmetric ionic fragment emission. For instance CO is preferred to be ionized by the laser field pointing from C to O along the molecular axis \cite{20,21,22,23,24,25}. The created molecular ion with biased orientation can subsequently be dissociated by the same asymmetric ultrashort laser pulse and may involve laser-induced coupling of electronic states, which in itself might result in an asymmetric ionic fragment emission \cite{26,27,28,29,30,addPRX}. We note that the controllable directional strong-field dissociative ionization was recently demonstrated in multiply charged states \cite{9,29,30}, polyatomic \cite{10,31} and hydrocarbon molecules \cite{32,33,34,35,35a}, and further in two-dimensional space \cite{36,37}. One essential aspect to thoroughly understand the directional dissociative ionization of complex molecules is to clearly distinguish individual contributions of the ionization and dissociation steps. It will also allow us to testify for the directional dissociation of a multi-electron system the role of laser-induced coupling of various electronic states, which rules the directional dissociation of the one-electron molecule H$_2^+$ \cite{17,18}. However, the coexistence of asymmetric ionization and asymmetric dissociation within a single femtosecond laser pulse blurs the contributions responsible for the ultimate directional emission of ionic fragments.

In this paper, using CO as a prototype, we conceived a strategy to disentangle the contributions from either step. As illustrated in Fig. \ref{fig1}, the single ionization created CO$^+$ cation by an elliptically polarized pump pulse, i.e. CO + n$h\nu_{\rm{pump}}$ $\rightarrow$ CO$^+$ + $e$, is dissociated by a time-delayed linearly polarized two-color pulse into C$^+$ and a neutral O atom, labeled as (C$^+$, O). The distinguished momentum distributions of electrons governed by the light polarizations allow us to identify the ionization and the dissociation induced by the pump and probe pulses. By coincidentally detecting the released electron and C$^+$, we can attribute the observed directional ejection of C$^+$ upon the bond breaking of CO$^+$ to the phase-dependent laser-induced coupling of various electronic states. The mechanism is numerically verified by solving the time-dependent Schr\"{o}dinger equation (TDSE).

\section{DISENTANGLING IONIZATION AND DISSOCIATION}

\begin{figure}
\includegraphics[width=0.95\columnwidth]{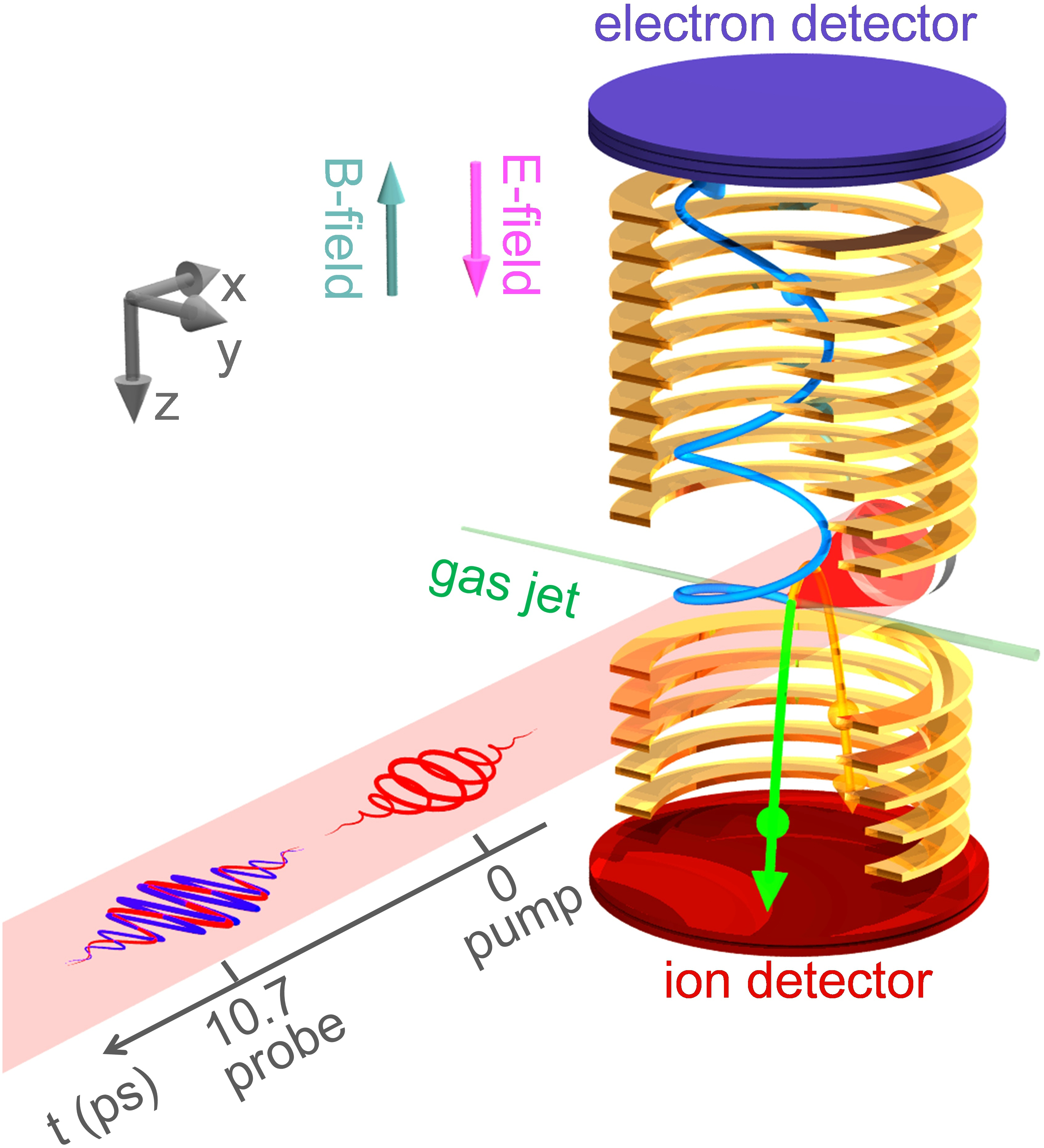}
\caption{(Color online) (a) Schematic diagram of the experimental apparatus. }
\label{fig1}
\end{figure}

The experiments were performed in an ultrahigh vacuum cold-target recoil ion momentum spectroscopy (COLTRIMS) setup \cite{38,39}, as illustrated in Fig. \ref{fig1}, driven by an elliptically polarized pump (ellipticity $\sim$ 0.8) and linearly polarized two-color probe femtosecond laser pulses. A femtosecond laser pulse from a Ti:sapphire multipass amplifier (25 fs, 795 nm, 10 kHz) was split into pump and probe pulses. The pump pulse was adjusted to be elliptically polarized in the \textit{y-z} plane with the major and minor axes along \textit{y}- and \textit{z}-axis, respectively. The two-color probe pulse was generated in a collinear scheme. Briefly, the \textit{z}-direction polarized fundamental-wave (FW) was frequency doubled in a 200 $\mu$m-thick $\beta$-barium borate (BBO) crystal to produce a second harmonic (SH) through the type-I phase matching. The polarization of the FW was rotated to be parallel to that of the SH along the \textit{y}-axis by using a dual-color wave plate. The time lag between the FW and SH pulses was compensated by a birefringent $\alpha$-BBO crystal. A pair of fused silica wedges were used to continuously vary the relative phase $\phi_{\rm{L}}$ between the FW and SH waves of the two-color pulse. The pump and probe pulses were collinearly combined using a beam splitter, which were afterwards sent into the vacuum chamber and focused onto the molecular beam using a concave silver mirror with a focusing length of \textit{f}=$75$ mm. The molecular beam was generated by supersonically co-expanding a mixture of 10\% CO and 90\% He through a 30 $\mu$m nozzle with a driving pressure of 1.5 bar. The intensities of the pump pulse, the FW and SH fields of the two-color probe pulse on the supersonic molecular beam of CO were measured to be 2.3$\times10^{14}$, 6.5$\times10^{13}$, and 8$\times10^{12}$ W/cm$^2$, respectively. To avoid the influence of impulsive molecular alignment (rotational period T$_{rot}$ = 8.64 ps for CO) by the pump pulse \cite{20,25}, the two-color probe pulse was time delayed by 10.7 ps to dissociate a randomly orientated CO$^+$ ensemble created by the pump pulse. In our measurement, the count rate on the electron detector was $\sim$ 0.24 electrons per laser shot with an ion-to-electron count ratio of $0.4:1$. The false coincidence was estimated to be\ $\sim$ 10 $\%$. To further suppress the electron-ion false coincidence for the single ionization dynamics, only events with just one detecting electron were selected for the data analysis.

\begin{figure}
\includegraphics[width=0.95\columnwidth]{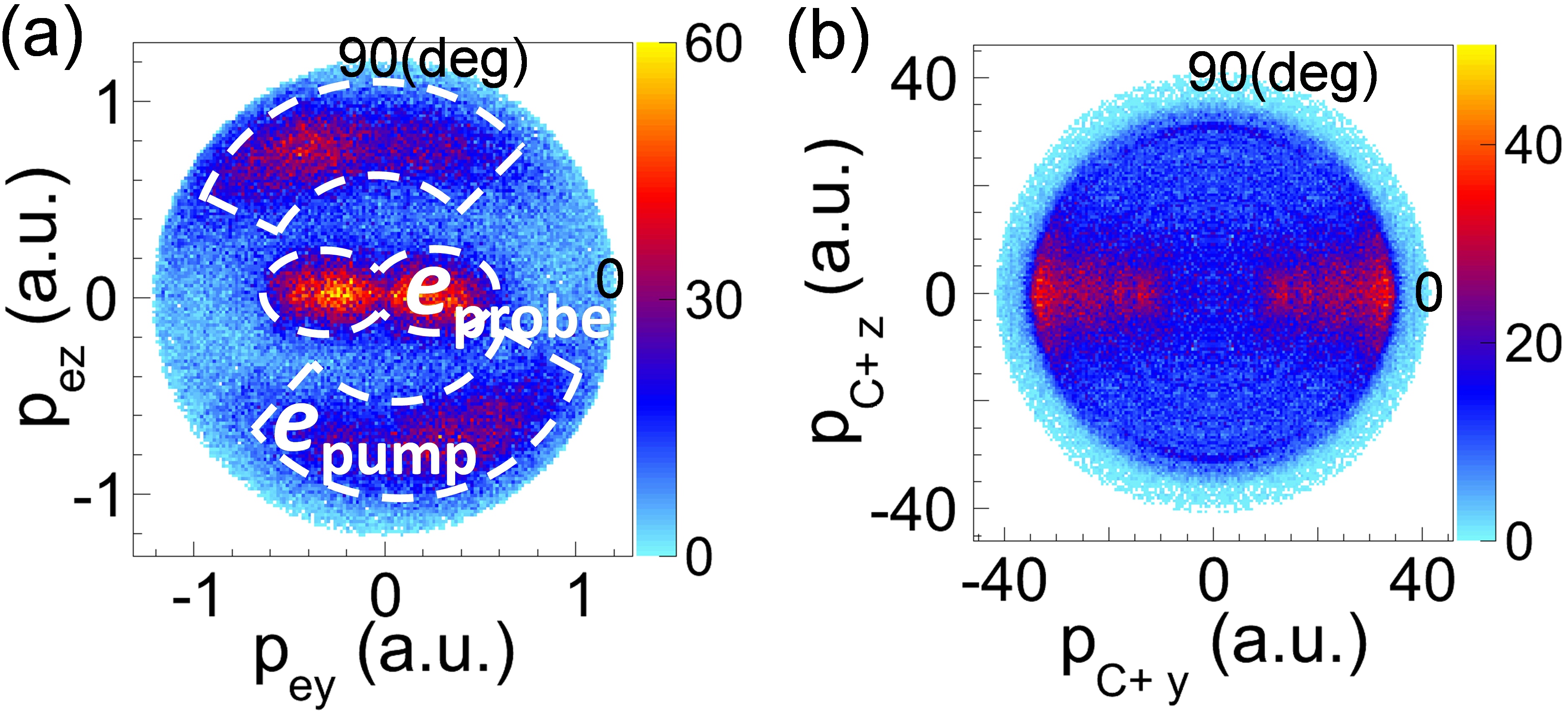}
\caption{(Color online) Momentum distributions of (a) emitted electrons measured in coincidence with (b) C$^+$ fragments of the (C$^+$, O) breakup channel.
}
\label{fig2}
\end{figure}

Figure \ref{fig2}(a) displays the momentum distribution of electrons freed by the pump and probe pulses measured in coincidence with the (C$^+$, O) breakup. The momentum distribution of C$^+$ from the (C$^+$, O) channel is shown in Fig. \ref{fig2}(b). The elliptically polarized pump pulse mostly liberates the electron along the major axis, which is afterwards angularly streaked to the minor axis of the polarization, i.e., the $e_{\rm{pump}}$ region in the dashed sectors in Fig. \ref{fig2}(a). The slight shift of the electron momentum distribution to the second and fourth quadrants is mainly due to the Coulomb potential effect of the ionic core on the departing electron \cite{41,42,43,44,45}. However, the electron freed by the linearly polarized probe pulse concentrates along the polarization direction, i.e., the $e_{\rm{probe}}$ region in the dashed ellipse in Fig. \ref{fig2}(a). Since only events with one detected electron are selected, the observed (C$^+$, O) breakup correlated to electrons in the $e_{\rm{probe}}$ region stands for both ionization and dissociation by the probe pulse. On the other hand, electrons in the $e_{\rm{pump}}$ region correspond to the ionization by the pump pulse, while the created CO$^+$ could be dissociated into the (C$^+$, O) pair either by the pump pulse itself or later on by the time-delayed probe pulse. We are mostly interested in the latter case, i.e. the ionization by the pump pulse and dissociation by the probe pulse. This scenario excludes the influence of molecular orientation-dependent field ionization and thus reveals the role of laser-induced coupling of various electronic states of the molecular ion on the directional dissociation.

\begin{figure}
\includegraphics[width=1.\columnwidth]{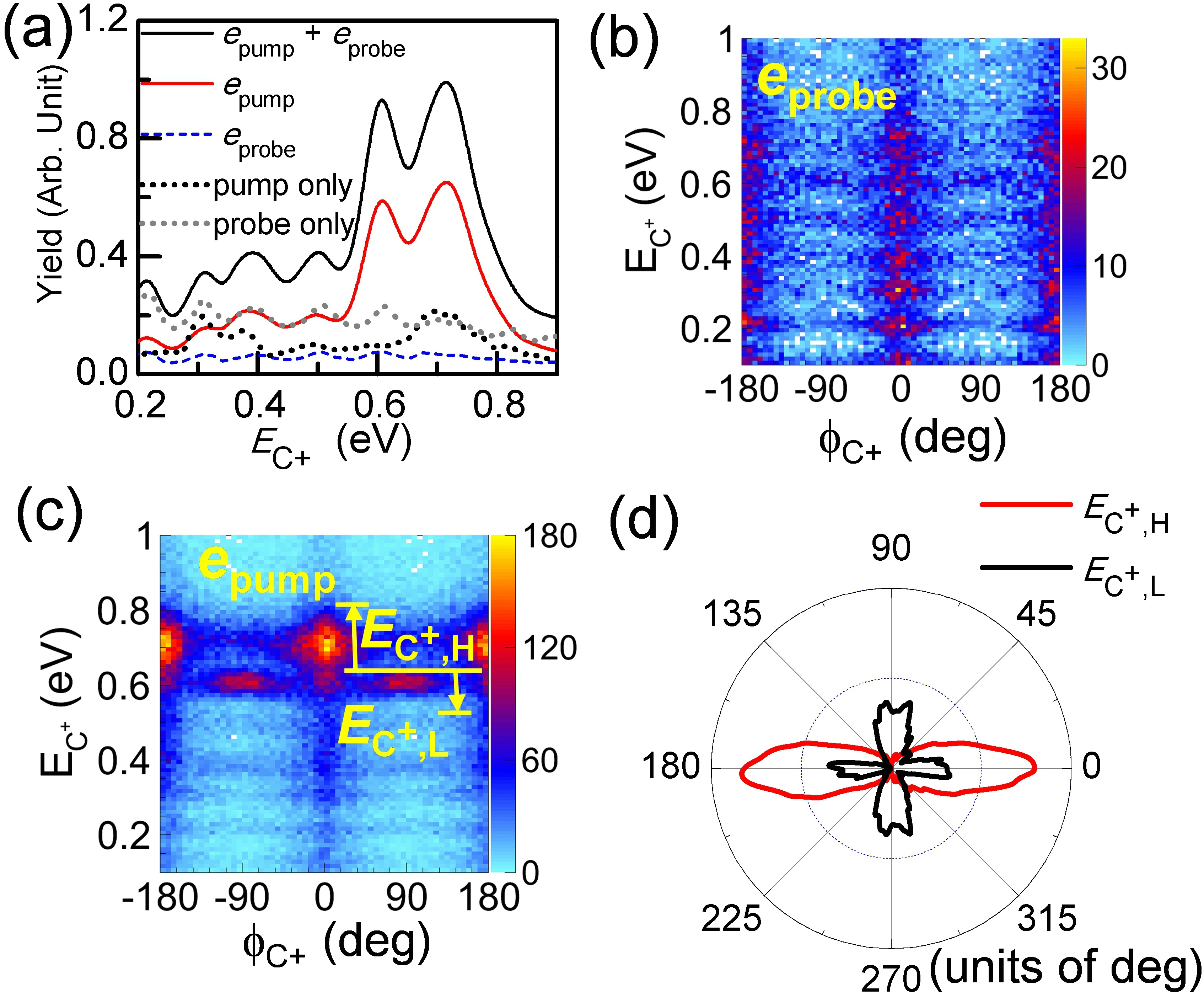}
\caption{(Color online) (a) KER spectra of C$^+$ of the (C$^+$, O) breakup channel measured in coincidence with electrons from the pump and probe pulses. See text for the descriptions of the curves. (b, c) $E_{\rm{C^+}}$-dependent angular distributions of C$^+$ of the (C$^+$, O) channel measured in coincidence with electrons in (b) $e_{\rm{probe}}$ and (c) $e_{\rm{pump}}$ regions depicted in Fig. \ref{fig2}(a). (d) Polar plots of the angular distribution of the emitted C$^+$ with KERs at $E_{\rm{C^+,L}}$ and $E_{\rm{C^+,H}}$ marked in (c).
}
\label{fig3}
\end{figure}

To extract the real pump-ionization probe-dissociation (C$^+$, O) breakup events, we further testify the kinetic energy release (KER) and angular distribution of the emitted C$^+$ fragments measured in coincidence with the electrons in the $e_{\rm{pump}}$ region. As displayed in Fig. \ref{fig3}(a), as compared to those by only the pump pulse (black dotted curve, legend ``pump only") or only the probe pulse (gray dotted curve, legend ``probe only"), the yield of the C$^+$ is significantly enhanced for $E_{\rm{C^+}}$ $>$ 0.57 eV when the probe pulse is sequentially applied following the pump pulse (black solid curve, legend ``$e_{\rm{pump}}$+$e_{\rm{probe}}$" correlated to all photoelectrons). The significant enhancement and the similar positions of the KER peaks indicate that (C$^+$, O) breakup is governed by the two-step process: the pump pulse singly ionizes CO, and the produced CO$^+$ is later on dissociated by the probe pulse. Such a two-step process is further confirmed by gating on the momentum distribution of the electron measured in coincidence with the ion fragments. As shown in Fig. \ref{fig3}(a), for electrons in the $e_{\rm{probe}}$ region (blue dashed curve, legend ``$e_{\rm{probe}}$"), the C$^+$ shows a similar KER distribution to that produced by the probe pulse only. The enhancement at $E_{\rm{C^+}}$ $>$ 0.57 eV is observed only for the (C$^+$, O) breakup when the electron is freed by the pump pulse and obtains momentum in the $e_{\rm{pump}}$ region (red solid curve, legend ``$e_{\rm{pump}}$"). The successive dissociation of the pump-created CO$^+$ by the time-delayed probe pulse leads to the enhanced KER distribution of C$^+$.

More interestingly, as displayed in Figs. \ref{fig3}(b) and \ref{fig3}(c), the emitted C$^+$ correlated to different electrons shows different KER-dependent angular distributions. Associated with electrons in the $e_{\rm{probe}}$ region, the C$^+$ mainly emits along the polarization direction of the linearly polarized two-color probe pulse, i.e. along $\phi_{\rm{C^+}} = 0\grad$ or $\pm 180\grad$. However, associated with electrons in the $e_{\rm{pump}}$ region, the angular distribution of the emitted C$^+$ strongly depends on $E_{\rm{C^+}}$. As shown in Fig. \ref{fig3}(c), C$^+$ emits mainly along $\phi_{\rm{C^+}} = 0\grad$ or $\pm 180\grad$ for $E_{\rm{C^+}}$ $<$ 0.57 eV, which is similar to the data in Fig. \ref{fig3}(b), and also similar to that by the pump pulse only (data not shown here). Thus we conclude that this part is produced by the pump pulse itself. Furthermore, a noticeable dissociation of orthogonally and parallel oriented molecules with the energy peaked at $E_{\rm{C^+}} = 0.6$ eV and 0.7 eV, respectively, is also observed. Figure \ref{fig3}(d) shows the corresponding angular distributions of the emitted C$^+$ for the high ($E_{\rm{C^+,H}}=$ 0.64-0.8 eV) and low ($E_{\rm{C^+,L}} =$ 0.57-0.64 eV) KER regions, respectively. The $E_{\rm{C^+}}$-dependent preferred dissociation of the CO$^+$ cation with the molecular axis parallel or orthogonal to the laser polarization indicates different dissociation dynamics.

\begin{figure}
\centerline{ \includegraphics[width=1\columnwidth]{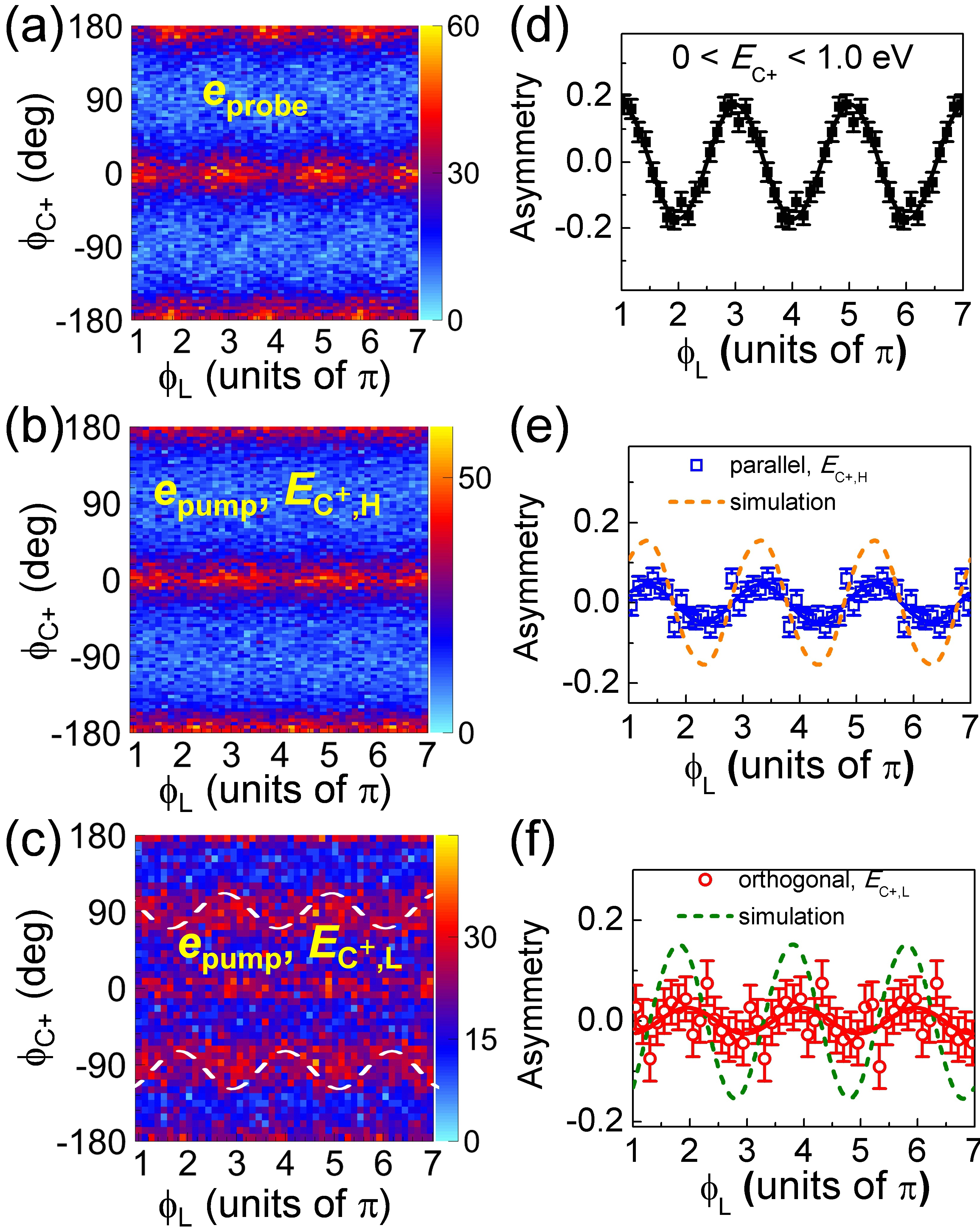}}
\caption{(Color online) (a--c) The $\phi_{\rm{L}}$-dependent yield of C$^+$ as a function of $\phi_{\rm{C^+}}$ at different KERs measured in coincidence with electrons in the (a) $e_{\rm{probe}}$ region, (b) $e_{\rm{pump}}$ region at $E_{\rm{C^+,H}}$ and (c) $e_{\rm{pump}}$ region at $E_{\rm{C^+,L}}$. (d)The corresponding asymmetries of the directional emission of C$^+$ versus $\phi_{\rm{L}}$. (e) The corresponding asymmetries of the directional emission of C$^+$ versus $\phi_{\rm{L}}$ (blue square) and the numerically simulated asymmetries when the molecule orients parallel (orange dashed curve) to the field polarization. (f) The corresponding asymmetries of the directional emission of C$^+$ versus $\phi_{\rm{L}}$ (blue square) and the numerically simulated asymmetries when the molecule orients orthogonal (olive dashed curve) to the field polarization.
}
\label{fig4}
\end{figure}

We will now address the essential question of whether or not the laser-induced coupling of various electronic states plays a crucial role in the directional dissociation of the multielectron system. We trace the directional emission of C$^+$ as a function of $\phi_{\rm{L}}$ of the two-color probe pulse. Figures \ref{fig4}(a)--4(c) display the $\phi_{\rm{L}}$-dependent yield of C$^+$ as a function of $\phi_{\rm{C^+}}$ at different KERs measured in coincidence with electrons in the $e_{\rm{pump}}$ or $e_{\rm{probe}}$ regions. Clear modulations of C$^+$ yield as functions of $\phi_{\rm{C^+}}$ and $\phi_{\rm{L}}$ are observed. To quantify the directional emission of C$^+$, we define the asymmetry parameter as $\beta(\phi_{\rm{L}}, \phi_{\rm{C^+}}) = [N(\phi_{\rm{L}}, \phi_{\rm{C^+}}) - N(\phi_{\rm{L}}+ \pi, \phi_{\rm{C^+}})] / [N(\phi_{\rm{L}}, \phi_{\rm{C^+}}) + N(\phi_{\rm{L}} + \pi, \phi_{\rm{C^+}})]$, where $N(\phi_{\rm{L}}, \phi_{\rm{C^+}})$ is the C$^+$ yield at emission angle $\phi_{\rm{C^+}}$ and phase $\phi_{\rm{L}}$ of the two-color probe pulse within certain energy ranges. The corresponding $\phi_{\rm{L}}$-dependent asymmetries are displayed in Figs. \ref{fig4}(d), \ref{fig4}(e) and \ref{fig4}(f). Here the statistical error bar of the asymmetry is calculated using the formula $2 \sqrt{a b / (a + b)^3}$, where $a=N(\phi_{\rm{L}}, \phi_{\rm{C^+}})$ and $b=N(\phi_{\rm{L}} + \pi, \phi_{\rm{C^+}})$, respectively \cite{PhDthesis}.  As shown in Figs. \ref{fig4}(a) and \ref{fig4}(d), where both ionization and dissociation are triggered by the probe pulse, C$^+$ is preferentially emitted to the direction opposite to the maxima of the two-color field, consistent with previous observations \cite{21,47}.

We now analyze the directional emission of C$^+$ in the pump-ionization probe-dissociation process. As shown in Figs. \ref{fig4}(b) and \ref{fig4}(e) (blue squares), directional emission of C$^+$ for $E_{\rm{C^+,H}}$ is clearly observed along the polarization direction of the two-color pulse. Note that there is no asymmetry of the molecular orientation in the initial ionization by the multicycle elliptically polarized pump pulse. The observed asymmetry in Fig. \ref{fig4}(e) (blue squares) should originate from the dissociation by the time-delayed two-color probe pulse. Since the electron has already been released by the elliptically pump pulse in the ionization step, recollisional excitation \cite{27,28,29,30} can be ruled out and laser-coupled transitions among various electronic states dominate the observed asymmetry depending on the relative phase of the two-color pulse. This is also consistent with previous observations \cite{28} that the emitted C$^+$ at KER lower than 1 eV is mainly due to the laser-induced coupling of the bound and repulsive electronic states. As compared to the asymmetry displayed in Fig. \ref{fig4}(d) where both the ionization and dissociation steps contribute the asymmetry for the C$^+$ emission, the $\phi_{\rm{L}}$-dependent asymmetry (blue squares) in Fig. \ref{fig4}(e) is much smaller and slightly phase shifted. In addition to molecules oriented along the light polarization, asymmetric dissociation of CO$^+$ is also observed for orthogonally oriented molecules around $\phi_{\rm{C^+}} = \pm 90\grad$ for C$^+$ at $E_{\rm{C^+,L}}$ as shown in Figs. \ref{fig4}(c) and \ref{fig4}(f) (red circles). As marked by the white dashed curve in Fig. \ref{fig4}(c), the asymmetry for the orthogonal molecule shown in Fig. \ref{fig4}(f) accounts for the preferred emission of C$^+$ diverging from $\phi_{\rm{C^+}} = \pm 90\grad$, i.e. C$^+$ is preferentially emitted to $\phi_{\rm{C^+}} = 90\grad- \delta_\phi ({\rm{or}} -90\grad+\delta_\phi)$ for $\phi_{\rm{L}} = 0$ as compared to $\phi_{\rm{C^+}} = 90\grad+\delta_\phi ({\rm{or}} -90\grad- \delta_\phi)$ for $\phi_{\rm{L}}={\uppi}$, where $\delta_\phi$ is a small angle. As shown in Figs. \ref{fig4}(e) and \ref{fig4}(f), the asymmetries for the parallel and orthogonally orientated molecules have different phases. We emphasize that the orthogonal orientation here only means that CO$^+$ has an orientation angle very close to but outside of $\pm 90\grad$. Strictly speaking, C$^+$ symmetrically emits with the exact angle $\pm90\grad$.
\\
\\
\section{NUMERICAL SIMULATION}

To explore how the directional emission of C$^+$ is built in the dissociation of CO$^+$, we numerically simulate the modelled TDSE (atomic units are used throughout unless indicated otherwise)

\begin{figure}
\includegraphics[width=0.95\columnwidth]{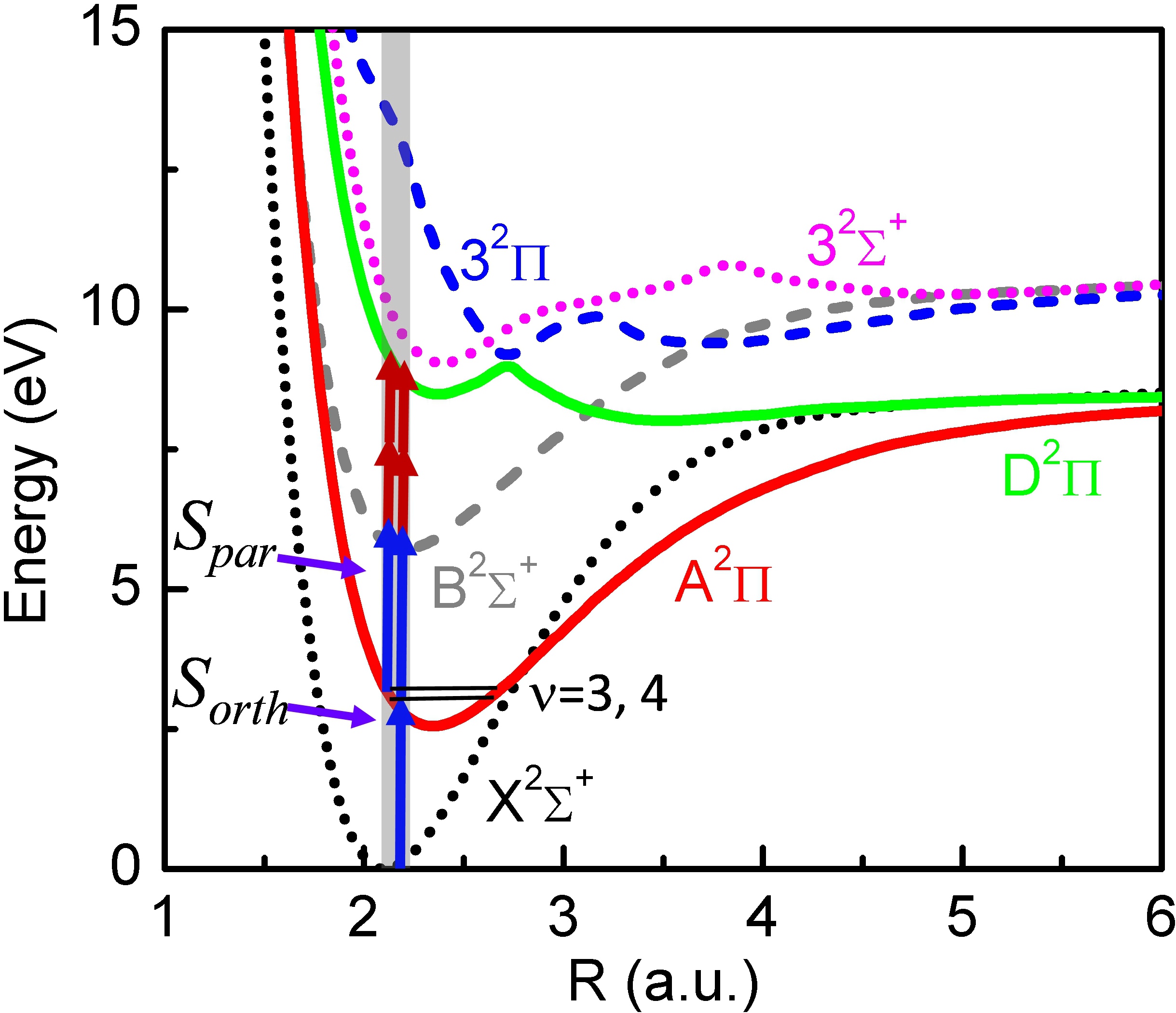}
\caption{(Color online) The potential energy surfaces of involved electronic states of C$^+$  calculated by MOLPRO, where the Franck-Condon ionization region of the ground state CO is indicated by the grayed bar. The red (FW photon) and blue (SH photon) arrows indicate the transition pathways for the dissociation of CO$^+$ with molecular axis parallel and orthogonal (labelled as \textit{S}$_{par}$ and \textit{S}$_{ort}$) to the polarization of the two-color field.}
\label{fig5}
\end{figure}

\begin{widetext}
\begin{equation}
i\frac{\partial}{\partial t}\left( {\begin{array}{*{20}c}
   \chi_1(R,t)   \\
   \chi_2(R,t)      \\
   \chi_3(R,t)      \\
   \vdots
\end{array}} \right) = \left( {\begin{array}{*{20}c}
   T_{nuc}+V_1(R) & \vec{\mu_{12}}\cdot\vec{E(t)} & \vec{\mu_{13}}\cdot\vec{E(t)} & \cdots \\
   \vec{\mu_{12}}\cdot\vec{E(t)} & T_{nuc}+V_2(R) & \vec{\mu_{23}}\cdot\vec{E(t)} & \cdots \\
   \vec{\mu_{13}}\cdot\vec{E(t)} & \vec{\mu_{23}}\cdot\vec{E(t)} & T_{nuc}+V_3(R) & \cdots \\
   \vdots & \vdots & \vdots & \ddots
\end{array}} \right)  \left( {\begin{array}{*{20}c}
   \chi_1(R,t)   \\
   \chi_2(R,t)      \\
   \chi_3(R,t)      \\
   \vdots
\end{array}} \right) ,
\label{eq:refname1}
\end{equation}
\end{widetext} where six electronic states are included to describe the dissociative dynamics, $\chi_{1\sim3}$ and $\chi_{4\sim6}$ are the associated nuclear wave packets for the three $\Sigma$ states and three $\Pi$ states shown in Fig. \ref{fig5} from bottom to up, respectively. T$_{nuc}$ is the nuclear kinetic energy operator, the potential energy curves $V{_i}(R)$ and the $R$-dependent dipole coupling matrix elements $\vec{\mu}_{ij}$ ($1 \leq i \leq 6, 1 \leq j \leq 6, i \neq j$) are calculated by the MOLPRO \cite{48} with the multi-reference configuration interaction method based on the aug-cc-pVQZ basis set \cite{49}. The reference configurations are all electronic configurations generated from [1$\upsigma^2$,2$\upsigma^2$,3$\upsigma^{0-2}$,4$\upsigma^{0-2}$,1$\uppi^{0-4}$,5$\upsigma^{0-2}$,2$\uppi^{0-4}$,6$\upsigma^{0-2}$,7$\upsigma^{0-1}$], and the calculations are performed under C$_{2v}$ symmetry \cite{49}. The two-color probe pulse $E(t)$ is written as

\begin{eqnarray}
  E(t) &=& E_1{\rm{cos}}(\omega_1t){\rm{exp}}[-4{\rm{ln}}{2\frac{(t-\frac{\tau_1}{2})^2}{\tau_1^2}}] \nonumber \\
  &+&E_2{\rm{cos}}(\omega_2(t-\Delta t)){\rm{exp}}[-4{\rm{ln}}{2\frac{(t-\Delta t-\frac{\tau_2}{2})^2}{\tau_2^2}}],
  \label{eq:refname2}
\end{eqnarray}
where the laser parameters are the same with that used in the experiment, $\Delta$t is the time delay. We used the Crank-Nicolson method to propagate the wave packets. The time and spatial steps are $\Delta$t = 0.1 a.u. and $\Delta$R = 0.02 a.u., and the simulation convergence has been tested by using denser time-spatial grids. The simulation box is big enough to hold all wave packets in all simulations.

\begin{figure}[htbp]
\centering
\includegraphics[width=0.8\linewidth]{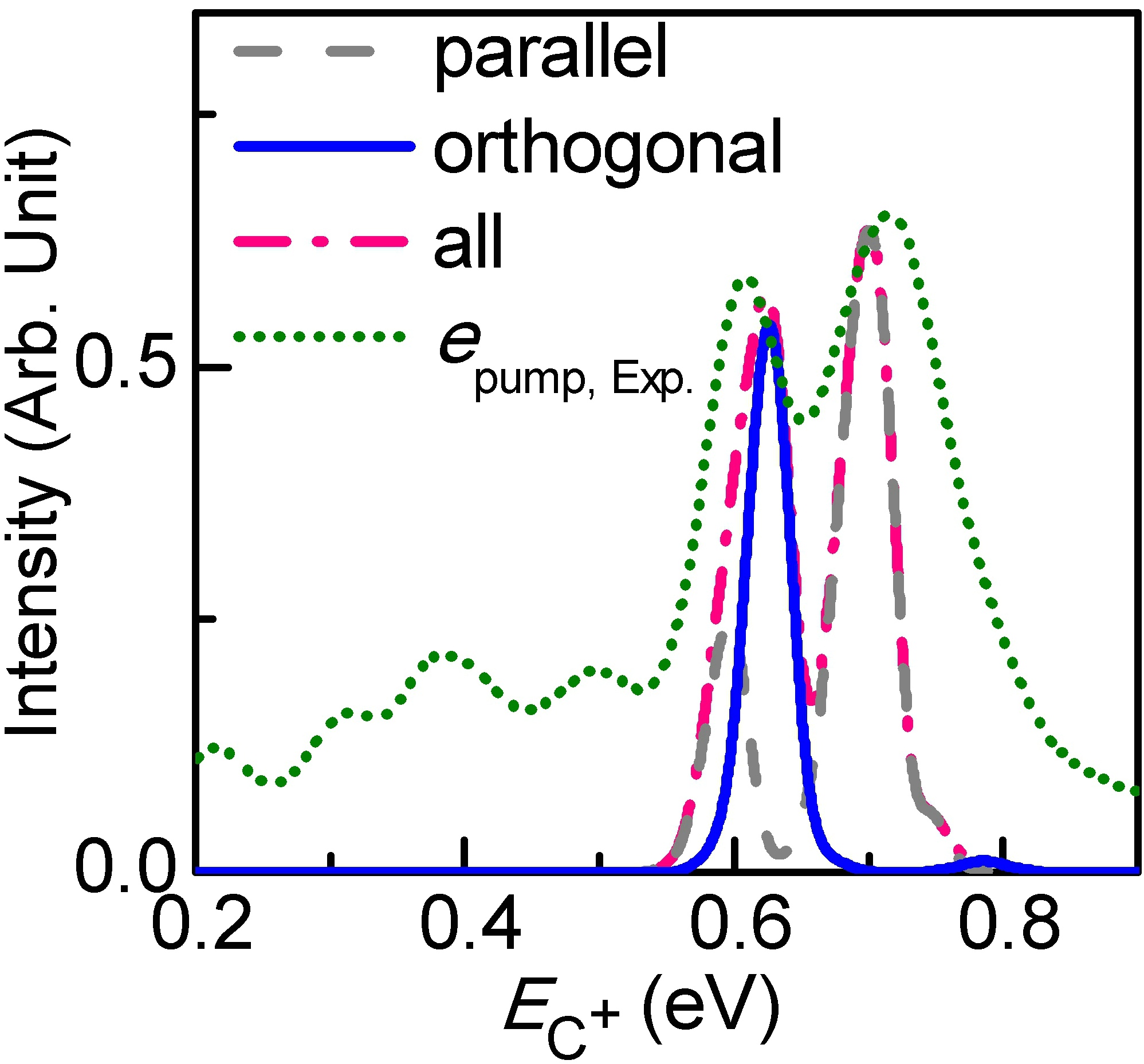}
\caption{(Color online) Simulated energy spectra of C$^+$ as the molecular axis is parallel (gray dashed curve) and orthogonal (blue solid curve) to the polarization of the two-color pulse. The addition of these two curves is presented by the pink dash-dotted curve, which is comparable with the experimental observations (olive dotted curve).}
\label{fig6}
\end{figure}

To reproduce the main observations of the experiment, we focused on two cases with the molecular axis parallel or orthogonal to the polarization of the two-color pulse. For C$^+$ emitted around the angle $\phi_{\rm{C^+}} = 0\grad$ or $180\grad$, the molecular axis of CO$^+$ is parallel to the polarization axis of the probe pulse when dissociation starts. We started from the vibrational states \textit{v}=3 or 4 of A$^2\Pi$ produced by two--$\omega_{pump}$-photon resonance excitation from the X$^2\Sigma^+$ state separately, which are obtained by the imaginary time propagation, and simulated Eq. (\ref{eq:refname1}) by only keeping all $^2\Pi$ states, and added the dissociative C$^+$ energy spectra incoherently after the interaction, i.e. the nuclear wave packet propagating on the A$^2\Pi$ curve undergoes a one-$\omega_{SH}$-photon and a two-$\omega_{FW}$-photon transition to the D$^2\Pi$ curve. Note that the simulation from the coherent sum of the vibrational states \textit{v}=3 or 4 gives almost the same result with the incoherent simulation since the dissociation fragments from \textit{v}=3 or 4 end with clearly different energies. The initial populations of these two vibrational states are assumed equally, and the final dissociation probabilities from these two vibrational states are adjusted by the relative weights obtained from experimental measurements. After the interaction with the probe pulse, we used the windows operator \cite{s4} to extract the energy spectra
\begin{equation}
\begin{split}
\scalebox{1.0}{$
P(E)=\sum\limits_{k}\widetilde{\left| \chi_k(E) \right|}^2,
$}
\end{split}
\label{eq:refname3}
\end{equation}
where $\widetilde{\chi_k(E)}=\dfrac{2}{\pi\delta_E}\langle\chi_k\left| R^+R \right|\chi_k\rangle$ and $R=\dfrac{\delta_E^2}{(E-H_0)^2+i\delta_E^2}$ with $E$ the total energy, $H_0$ the field free Hamiltonian and $\delta_E$ associated with the energy resolution. The simulated C$^+$ energy spectra are shown in Fig. \ref{fig6} by the gray dashed curve, which qualitatively agrees well with the experimental observations (olive dotted curve). The calculated asymmetry parameter of C$^+$ is shown in Fig. \ref{fig4}(e) (orange dashed curve).The low energy peaks in the range 0.2-0.5 eV shown in Fig. \ref{fig3}(a) are contributed by other vibrational states of A$^2\Pi$, which are vertically populated when the HOMO-1 electron in CO is directly removed by the pump pulse.

For C$^+$ emitted around the angle $\phi_{C^+}$= 90\grad, we aligned CO$^+$ with an angle of $80\grad$ in calculations. We tested that the conclusion does not change substantially if the aligned angle varies between $75\grad$ and $85\grad$. In this case, all transitions between these six electronic states are allowed. We chose \textit{v} =0 of X$^2\Sigma^+$ state of CO$^+$ as the initial state, which is also almost the ground nuclear state of CO. The component of the laser field perpendicular to the molecular axis triggers the transition between $^2\Sigma^+$ and $^2\Pi$, and the parallel component of the laser field to the molecular axis induces the transition between all $^2\Pi$ states, i.e. the nuclear wave packet propagating on the X$^2\Sigma^+$ curve undergoes one-$\omega_{SH}$-photon transition to A$^2\Pi$, then undergoes a one-$\omega_{SH}$-photon and a two-$\omega_{FW}$-photon transition to the D$^2\Pi$ curve. We followed the same steps as calculating C$^+$ emitted along the angle $\phi_{C^+}= 0\grad$ or $180\grad$, and show the calculated energy spectrum and asymmetry in Fig. \ref{fig6} (blue solid curve) and Fig. \ref{fig4}(f) (olive dashed curve), respectively.
To fully reproduce the angular distribution as shown in Figs. \ref{fig3}(b) and \ref{fig3}(c), a numerical model which describes the nuclear wave packet in two-dimensional space is to be developed. Nevertheless, the good qualitative agreement of the energy spectra and asymmetries between experimental measurements and theoretical calculations confirms that the laser coupling between different electronic states dominates the asymmetric C$^+$ emission.
\\
\\
\section{CONCLUSION}

In summary, taking CO as a prototype, a straightforward and robust strategy to disentangling the coexisting ionization and dissociation contributions to the directional dissociative ionization of a multielectron molecule is demonstrated. Our quantum simulations confirm that the laser-coupling among different electronic states in CO$^+$ plays an important role for the directional C$^+$ fragment emission. Depending on the KER of the nuclear fragments, directional dissociation of CO$^+$ oriented orthogonally to the light polarization is observed as a function of the relative phase of the two-color ultrashort laser pulse. The strong-field dissociative ionization of molecules is complex where many effects may be involved and entangled with each other. Our experimental technique disentangles the dissociative ionization into the ionization and dissociation steps, providing a powerful tool to investigate even more complex molecular reactions in strong laser fields.

\section*{ACKNOWLEDGMENTS}
This work is supported by the National Natural Science Fund of China (Grant Nos. 11425416, 11374103, 11322438, 11574205, 11434005), and the Program of Introducing Talents of Discipline to Universities (Grant No. B12024). M.F.K. acknowledges support by the German Research Foundation (DFG) via LMUexcellent and the cluster of excellence ``Munich Centre for Advanced Photonics" and by the European Union (EU) through the ERC grant ATTOCO (no. 307203).


\begin{thebibliography}{20}
\bibitem{1}
M. F. Kling, Ch. Siedschlag, A. J. Verhoef, J. I. Khan, M. Schultze, Th. Uphues, Y. Ni, M. Uiberacker, M. Drescher, F. Krausz, and M. J. J. Vrakking, Science \textbf{312}, 246 (2006).
\bibitem{2}
M. Kremer, B. Fischer, B. Feuerstein, V. L. B. de Jesus, V. Sharma, C. Hofrichter, A. Rudenko, U. Thumm, C. D. Schr\"{o}ter, R. Moshammer, and J. Ullrich, Phys. Rev. Lett. \textbf{103}, 213003 (2009).
\bibitem{3}
B. Fischer, M. Kremer, T. Pfeifer, B. Feuerstein, V. Sharma, U. Thumm, C. D. Schr\"{o}ter, R. Moshammer, and J. Ullrich, Phys. Rev. Lett. \textbf{105}, 223001 (2010).
\bibitem{4}
J. McKenna, F. Anis, A. M. Sayler, B. Gaire, N. G. Johnson, E. Parke, K. D. Carnes, B. D. Esry, and I. Ben-Itzhak, Phys. Rev. A \textbf{85}, 023405 (2012).
\bibitem{5}
H. Xu, T. Xu, F. He, D. Kielpinski, R. Sang, and I. V. Litvinyuk, Phys. Rev. A \textbf{89}, 041403(R) (2014).
\bibitem{6}
B. Sheehy, B. Walker, and L. F. DiMauro, Phys. Rev. Lett. \textbf{74}, 4799 (1995).
\bibitem{7}
H. Ohmura, N. Saito, and M. Tachiya, Phys. Rev. Lett. \textbf{96}, 173001 (2006).
\bibitem{8}
D. Ray, F. He, S. De, W. Cao, H. Mashiko, P. Ranitovic, K. P. Singh, I. Znakovskaya, U. Thumm, G. G. Paulus, M. F. Kling, I. V. Litvinyuk, and C. L. Cocke, Phys. Rev. Lett. \textbf{103}, 223201 (2009).
\bibitem{9}
K. J. Betsch, D. W. Pinkham, and R. R. Jones, Phys. Rev. Lett. \textbf{105}, 223002 (2010).
\bibitem{10}
N. Kotsina, S. Kaziannis, S. Danakas and C. Kosmidis, J. Chem. Phys. \textbf{139}, 104313 (2013).
\bibitem{11}
A. D. Bandrauk, S. Chelkowski, and H. S. Nguyen, Int. J. Quantum Chem. \textbf{100}, 834 (2004).
\bibitem{12}
X. M. Tong and C. D. Lin, Phys. Rev. Lett. \textbf{98}, 123002 (2007).
\bibitem{13}
V. Roudnev and B. D. Esry, Phys. Rev. Lett. \textbf{99}, 220406 (2007).
\bibitem{14}
F. Kelkensberg, G. Sansone, M. Y. Ivanov, and M. Vrakking, Phys. Chem. Chem. Phys. \textbf{13}, 8647 (2011).
\bibitem{15}
J. Wu, M. Magrakvelidze, L. Ph. H. Schmidt, M. Kunitski, T. Pfeifer, M. Sch\"{o}ffler, M. Pitzer, M. Richter, S. Voss, H.
Sann, H. Kim, J. Lower, T. Jahnke, A. Czasch, U. Thumm, and R. D\"{o}rner, Nat. Commun. \textbf{4}, 2177 (2013).
\bibitem{16}
A. D. Bandrauk and S. Chelkowski, Phys. Rev. Lett. \textbf{84}, 3562 (2000).
\bibitem{17}
N. G. Kling, K. J. Betsch, M. Zohrabi, S. Zeng, F. Anis, U. Ablikim, B. Jochim, Z. Wang, M. K\"{u}bel, M. F. Kling, K. D. Carnes, B. D. Esry, and I. Ben-Itzhak, Phys. Rev. Lett. \textbf{111}, 163004 (2013).
\bibitem{18}
T. Rathje, A. M. Sayler, S. Zeng, P. Wustelt, H. Figger, B. D. Esry, and G. G. Paulus, Phys. Rev. Lett. \textbf{111}, 093002 (2013).
\bibitem{19}
Z. Wang, K. Liu, P. Lan and P. Lu, J. Phys. B \textbf{48}, 015601 (2015).
\bibitem{20}
S. De, I. Znakovskaya, D. Ray, F. Anis, Nora G. Johnson, I. A. Bocharova, M. Magrakvelidze, B. D. Esry, C. L. Cocke, I. V. Litvinyuk, and M. F. Kling, Phys. Rev. Lett. \textbf{103}, 153002 (2009); 112, 159902 (2014).
\bibitem{21}
H. Li, D. Ray, S. De, I. Znakovskaya, W. Cao, G. Laurent, Z. Wang, M. F. Kling, A. T. Le, and C. L. Cocke, Phys. Rev. A \textbf{84}, 043429 (2011).
\bibitem{22}
 J. Wu, L. Ph. H. Schmidt, M. Kunitski, M. Meckel, S. Voss, H. Sann, H. Kim, T. Jahnke, A. Czasch, and R. D\"{o}rner, Phys. Rev. Lett. \textbf{108}, 183001 (2012).
\bibitem{23}
 E. Frumker, C. T. Hebeisen, N. Kajumba, J. B. Bertrand, H. J. W\"{o}rner, M. Spanner, D. M. Villeneuve, A. Naumov, and P. B. Corkum, Phys. Rev. Lett. \textbf{109}, 113901 (2012).
\bibitem{24}
B. Zhang, J. Yuan, and Z. Zhao, Phys. Rev. Lett. \textbf{111}, 163001 (2013).
\bibitem{25}
I. Znakovskaya, M. Spanner, S. De, H. Li, D. Ray, P. Corkum, I. V. Litvinyuk, C. L. Cocke, and M. F. Kling, Phys. Rev. Lett. \textbf{112}, 113005 (2014).
\bibitem{26}
A. S. Alnaser, C. M. Maharjan, X. M. Tong, B. Ulrich, P. Ranitovic, B. Shan, Z. Chang, C. D. Lin, C. L. Cocke, and I. V. Litvinyuk, Phys. Rev. A \textbf{71}, 031403(R) (2005).
\bibitem{27}
I. Znakovskaya, P. von den Hoff, S. Zherebtsov, A. Wirth, O. Herrwerth, M. J. J. Vrakking, R. de Vivie-Riedle, and M. F. Kling, Phys. Rev. Lett. \textbf{103}, 103002 (2009).
\bibitem{28}
P. von den Hoff, I. Znakovskaya, M.F. Kling, R. de Vivie-Riedle, Chem. Phys. \textbf{366}, 139 (2009).
\bibitem{29}
Y. Liu, X. Liu, Y. Deng, C. Wu, H. Jiang, and Q. Gong, Phys. Rev. Lett. \textbf{106}, 073004 (2011).
\bibitem{30}
K. J. Betsch, Nora G. Johnson, B. Bergues, M. K\"{u}bel, O. Herrwerth, A. Senftleben, I. Ben-Itzhak, G. G. Paulus, R. Moshammer, J. Ullrich, M. F. Kling, and R. R. Jones, Phys. Rev. A \textbf{86}, 063403 (2012). 

\bibitem{addPRX}
A. Trabattoni, M. Klinker, J. Gonz\'{a}lez-V\'{a}zquez, C. Liu, G. Sansone, R. Linguerri, M. Hochlaf, J. Klei, M. J. J. Vrakking, F. Mart\'{i}n, M. Nisoli, and F. Calegari, Phys. Rev. X \textbf{5}, 041053 (2015).

\bibitem{31}
I. Bocharova, R. Karimi, E. F. Penka, J. Brichta, P. Lassonde, X. Fu, J. Kieffer, A. D. Bandrauk, I. Litvinyuk, J. Sanderson, and F. L\'{e}gar\'{e}, Phys. Rev. Lett. \textbf{107}, 063201 (2011).
\bibitem{32}
X. Xie, K. Doblhoff-Dier, S. Roither, M. S. Sch\"{o}ffler, D. Kartashov, H. Xu, T. Rathje, G. G. Paulus, A. Baltu\v{s}ka, S. Gr\"{a}fe, and M. Kitzler, Phys. Rev. Lett. \textbf{109}, 243001 (2012).
\bibitem{33}
X. Xie, K. Doblhoff-Dier, H. Xu, S. Roither, M. S. Sch\"{o}ffler, D. Kartashov, S. Erattupuzha, T. Rathje, G. G. Paulus, K. Yamanouchi, A. Baltu\v{s}ka, S. Gr\"{a}fe, and M. Kitzler, Phys. Rev. Lett. \textbf{112}, 163003 (2014).
\bibitem{34}
X. Gong, Q. Song, Q. Ji, H. Pan, J. Ding, J. Wu, and H. Zeng, Phys. Rev. Lett. \textbf{112}, 243001 (2014).
\bibitem{35}
Q. Song, X. Gong, Q. Ji, K. Lin, H. Pan, J. Ding, H. Zeng, and J. Wu, J. Phys. B: At., Mol. Opt. Phys. \textbf{48}, 094007 (2015).
\bibitem{35a}
M. K\"ubel, R. Siemering, C. Burger, Nora G. Kling, H. Li, A. S. Alnaser, B. Bergues, S. Zherebtsov, A. M. Azzeer, I. Ben-Itzhak, R. Moshammer, R. de Vivie-Riedle, and M. F. Kling, Phys. Rev. Lett. \textbf{116}, 193001 (2016).
\bibitem{36}
X. Gong, P. He, Q. Song, Q. Ji, H. Pan, J. Ding, F. He, H. Zeng, and J. Wu, Phys. Rev. Lett. \textbf{113}, 203001 (2014).
\bibitem{37}
K. Lin, X. Gong, Q. Song, Q. Ji, W. Zhang, J. Ma, P. Lu, H. Pan, J. Ding, H. Zeng, and J. Wu, J. Phys. B: At., Mol. Opt. Phys. \textbf{49}, 025603 (2016).
\bibitem{38}
R. D\"{o}rner, V. Mergel, O. Jagutzki, L. Spielberger, J. Ullrich, R. Moshammer, and H. Schmidt-B\"{o}cking, Phys. Rep. \textbf{330}, 95 (2000).
\bibitem{39}
J. Ullrich, R. Moshammer, A. Dorn, R. D\"{o}rner, L. Ph. H. Schmidt, and H. Schmidt-B\"{o}cking, Rep. Prog. Phys. \textbf{66}, 1463 (2003).
\bibitem{41}
G. G. Paulus, F. Grasbon, A. Dreischuh, H. Walther, R. Kopold, and W. Becker, Phys. Rev. Lett. \textbf{84}, 3791 (2000).
\bibitem{42}
A. N. Pfeiffer, C. Cirelli, A. S. Landsman, M. Smolarski, D. Dimitrovski, L. B. Madsen, and U. Keller, Phys. Rev. Lett. \textbf{109}, 083002 (2012).
\bibitem{43}
D. Shafir, H. Soifer, C. Vozzi, A. S. Johnson, A. Hartung, Z. Dube, D. M. Villeneuve, P. B. Corkum, N. Dudovich, and A. Staudte, Phys. Rev. Lett. \textbf{111}, 023005 (2013).
\bibitem{44}
M. Li, Y. Liu, H. Liu, Q. Ning, L. Fu, J. Liu, Y. Deng, C. Wu, L. Peng, and Q. Gong, Phys. Rev. Lett. \textbf{111}, 023006 (2013).
\bibitem{45}
P. He, C. Ruiz, and F. He, Phys. Rev. A \textbf{91}, 063413 (2015).
\bibitem{PhDthesis}
M. Meckel, Ph.D. thesis, Goethe University Frankfurt, 2011.
\bibitem{47}
J. Wu, A. Vredenborg, L. Ph. H. Schmidt, T. Jahnke, A. Czasch, and R. D\"{o}rner, Phys. Rev. A \textbf{87}, 023406 (2013).
\bibitem{48}
H.-J. Werner, P. J. Knowles, G. Knizia, F. R. Manby, M. Sch\"utz et al., MOLPRO, version 2015.1 (2015).
\bibitem{49}
K. Okada and S. Iwata, J. Chem. Phys. \textbf{112}, 1804 (2000).
\bibitem{s4}
K. J. Schafer and K.C. Kulander, Phys. Rev. A \textbf{42}, 5794 (1990).
\end{thebibliography}
\end{document}